# COMPARATIVE ANALYSIS OF THE ACCURACY OF THE DISTANCE TO THE OBSERVED OBJECT FOR GEOMETRIC METHODS.

S.A. Pyunninen

*North-West State Technical University,
St. Petersburg*

## ABSTRACT

The article presents a comparative analysis of the accuracy of the distance to the observed object for geometric methods in noisy observations of bearings-only information.

Algorithms for determining the parameters of the trajectories of moving objects are widely used in control systems and navigation systems, with one of the most difficult tasks is to determine the trajectory parameters for passive goniometric monitoring a moving object under the action of perturbations.

Conditions of the problem imply that the observer and the object of observation (target) is moving in two dimensions on a smooth trajectory. In the discrete time the observer measures the angle of the bearing to the target. Angle measurement is performed with error representing a white noise. Coordinates of an observer at each discrete point in time are assumed to be known. Must be based on observations to recover the trajectory of target with a specified accuracy.

In the mathematical sense, this task is the task of restoring function to a limited number of observation data. Feature of the problem is that the observation data at each discrete moment of time sufficient to obtain a unique solution.

Currently known methods for calculating the number of geometric coordinates and motion parameters of objects (KPDO) based on using only the goniometric information, with the greatest application is the method of N -Bearing [1]. A common feature of these methods is that they are applicable only for uniform rectilinear motion of the observed object, whereas in the wild movement of the observed object, as a rule, nonlinear.

A number of studies [2,3], the author proposed a method N Polynomials, which allows to determine the parameters of the trajectory in the case of nonlinear motion of the observed object. The basis of this method is an approximation of the object's trajectory observing the parametric functions of the coordinates of time, given as a linear combination of orthogonal polynomials [4]. As a basic system of polynomials can be used Chebyshev polynomials of the 1 st and 2 nd kind, as well as Legendre polynomials.

Distinctive features of the proposed method is:
1. opportunity to determine the trajectory nonlinear moving target;
2. ensuring continuity solution for any parameters of maneuvering targets;
3. separate mathematical processing of simulation results and an error estimate for each of the coordinates;
4. minimizing the approximation error of the trajectory through the use of orthogonal polynomials;
5. minimize errors of observation through the use of processing raw data by the method of least squares;

6. ease of obtaining the parameters of the trajectory at any point in time calculated by differentiation of the simulated functions.

Results compare the accuracy of determining the parameters of the trajectories nonlinear moving target for various models of movement by the method N Polynomials, and the method N -Bearing are shown in figures 1-3.

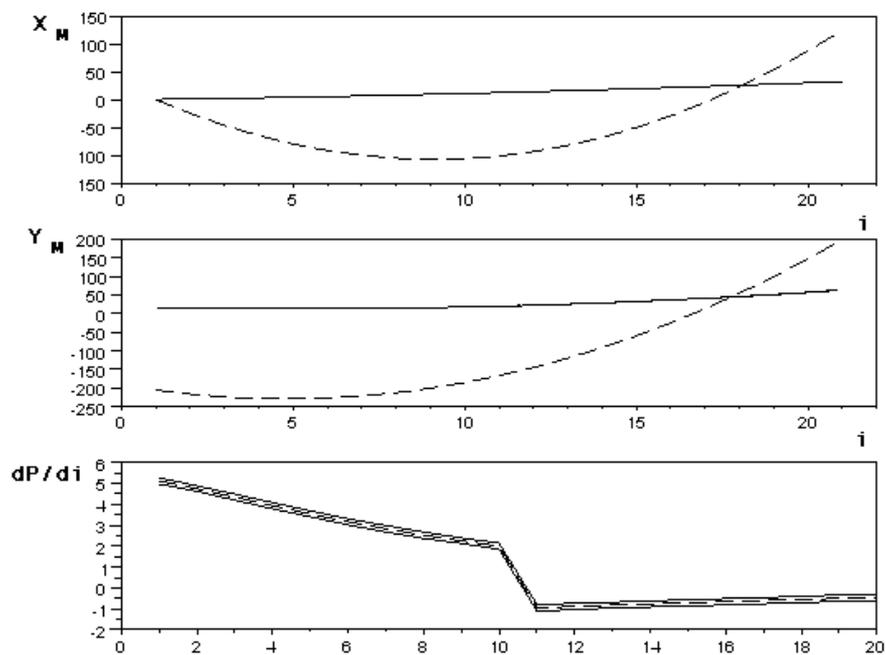

Figure 1. Errors in determining the origin for purposes of uniformlyac celerated motion on Big VIP

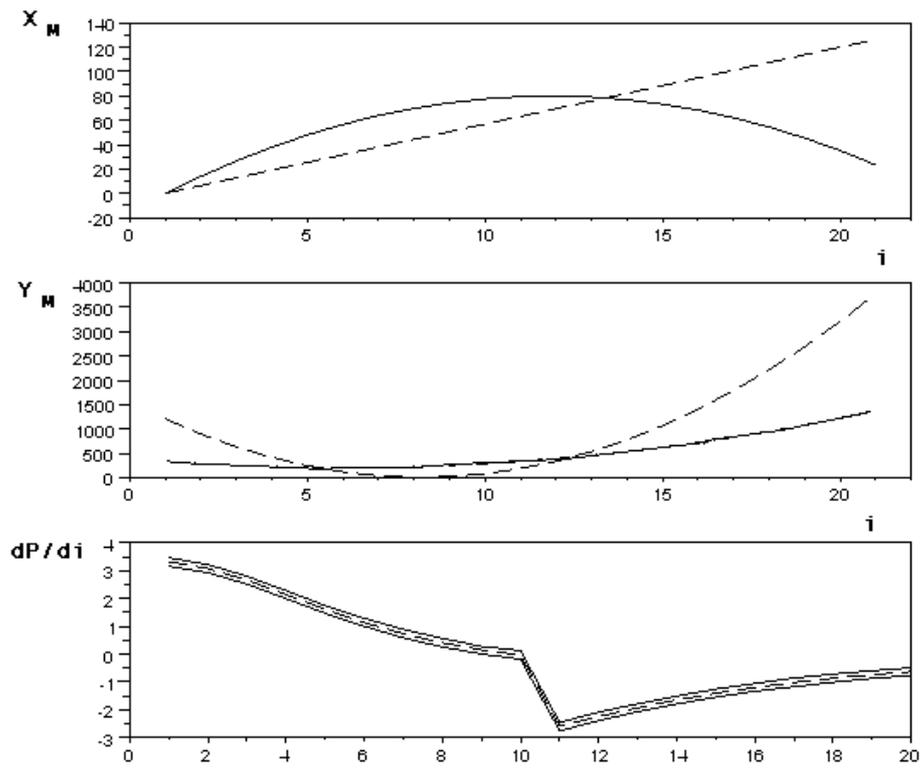

Figure 2. Errors in determining the origin for purposes of uniformly accelerated motion in a small VIP

Figure 3. Errors in determining the origin for moving target in a parabola on average VIP

Thus, use of N Polynomials to determine the parameters of the trajectory of a moving object in the nonlinear terms of noise, is preferred because it allows much more accurate solution of the problem and ensure the continuity of the solution.